\documentclass{ws-procs9x6-cpt16}
\begin{document}

\newcommand{\refeq}[1]{(\ref{#1})}
\def\etal {{\it et al.}}

\title{Positronium and Muonium 1S-2S Laser Spectroscopy\\ 
as a Probe for the Standard-Model Extension}

\author{P.\ Crivelli and G.\ Wichmann}

\address{Institute for Particle Physics, ETH Zurich, 
8093 Zurich, Switzerland}

\begin{abstract}
We present the status and the prospects of the ongoing efforts to improve the measurements of the 1S-2S transition in positronium and muonium. These purely leptonic systems represent ideal objects to test bound-state QED to high precision and the Standard-Model Extension.
\end{abstract}

\bodymatter

\section{Introduction}
Positronium ($e^+e^-$, Ps) and muonium ($\mu^+e^-$, M) being purely leptonic atomic systems are ideal to study bound-state QED free of finite size effects with hadronic contributions strongly suppressed compared to hydrogen.\cite{savely} They are also used for searches of new physics,\cite{Willmann,oPsInv} tests of fundamental symmetries,\cite{asaiPRL2010} and found their application in material science (see, e.g., Refs.\ \refcite{reviewMuAppl,reviewPsAppl}). In this contribution we will review the status and the prospects of the determination of the 1S-2S transition frequencies for Ps and M which are sensitive probes for the Standard-Model Extension (SME).\cite{Kostelecky} The 1S-2S transition was first observed for Ps in 1982\cite{ChuMills1982} and a few years later (1988) for M.\cite{ChuMills1988} In these two measurements positrons and positive muons were implanted in silica powders in which, by capturing an electron, ground state Ps or M were produced. A fraction of these atoms diffused to the silica surface and were emitted into vacuum where they were excited with pulsed lasers to the 2S state.
Subsequent measurements followed in 1984 for positronium\cite{ChuMills1984} with a more efficient target for Ps production made of Al(111) and in 1994 for muonium\cite{Maas1994} with a higher muon flux (almost two orders of magnitude larger). These upgrades allowed reaching uncertainties of 12 ppb and 21 ppb, respectively. 
For Ps, the use of a CW laser allowed a further improvement in the determination of the transition frequency at a level of 2.4 ppb in 1993.\cite{Fee1993} For M the best current measurement (4 ppb) was performed  in the year 2000 with a refined laser chirp control and the use of long pulses to reduce the time-of-flight broadening\cite{Meyer2000} by increasing the interaction time. 
The results of those measurements are in a good agreement with the QED calculations.\cite{savely}
Both experiments are statistically limited and would profit from new sources with a larger flux of colder Ps and M atoms. This can be achieved by more intense/brighter primary beams and/or improved conversion targets.\cite{Jungmann2006} In particular slower atoms would be a great advantage to reduce systematic effects and increase the signal rate because of the extended interaction time with the laser (the excitation probability is proportional to $t^2$).

\section{Status of positronium 1S-2S measurement}
For positronium, bound-state QED calculations reached a level of 0.5 ppb.\cite{QEDPs} A new measurement is ongoing at ETH Zurich\cite{hype15} and at UC Riverside \cite{mills2016} to improve the experimental accuracy to this level. New targets for stable Ps production were developed\cite{APL} and preliminary results were obtained detecting the annihilation of Ps in the 2S states.\cite{hype15} Those prompted the need for improved S/N ratio with the use of a buffer gas trap.\cite{Surko} Efficient extraction of the pulsed beam\cite{Cooke2015b} to a field-free region was achieved reducing the systematic from electromagnetic fields such as DC and motional Stark and Zeeman shifts. Excitation in Rydberg states with subsequent detection via field ionization 
will allow correcting for the second-order Doppler shift expected to be the main systematic effect of the experiment caused by the very light mass of Ps resulting in a high velocity ($10^5$ m/s) even at room temperature. The used positron bunches for Ps production have a narrow time window of $1\,\text{ns}$ and the Ps is emitted from porous silica with a well defined velocity determined by the ground state in the pores.\cite{oPsTOF,CassidyPRA2010} A time-of-flight measurement of the 2S-excited Ps atoms will be performed by their detection at a known distance from 
a plate in which field Ps atoms will be ionized. With this it is aimed to obtain the mean emission velocity to $\leq 4\%$ by comparison to simulated time spectra resulting in an uncertainty at a level of $100\,\text{kHz}$. An accuracy of 0.5 ppb seems thus in reach with the available Ps targets. Different schemes to produce colder Ps have been proposed including laser cooling (see, e.g., Ref.\ \refcite{crivelli2014}) and Stark deceleration of Rydberg Ps atoms\cite{Hogan2016} pointing to further possible improvements. 
\section{Status of muonium 1S-2S measurement}
Recent advances in the production of M into vacuum\cite{Antognini2012} and spatial confinement of M\cite{MuConf} enable CW spectroscopy with current UV technology and with the existing low-energy muon beam line at PSI.\cite{LEM} This will result in a narrower line (about 1 MHz) compared to pulsed lasers (20 MHz dominated by laser chirps). The main systematic effect due to the residual first-order Doppler shift will be at a negligible level due to the use of an enhancement cavity as in hydrogen spectroscopy.\cite{MPQ} A much higher degree of collinearity of the counter-propagating beams will be granted with this, allowing for a measurement of the 1S-2S transition frequency at the 0.2 ppb level (a factor of 20 improvement over Meyer in 2000\cite{Meyer2000}). This results in a test of bound-state QED with an actual theoretical uncertainty of 0.4 ppb.\cite{pachucki, karshenboim} It will also provide the best verification of charge equality in the first two generations of particles and it will improve the determination of the muon mass at the 40 ppb level (a factor 3 better than extracted from the hyperfine splitting measurement\cite{LiuHFS}). This experiment will be statistically limited, therefore,  ongoing efforts to develop a high-flux and high-brightness slow-muon beam line\cite{MuCool,Andreas, Strasser:2014dsa} promise even higher accuracy.

\section{Conclusions}
New improved measurements of the 1S-2S transition in positronium and muonium at a level of 0.5 ppb are ongoing and the results are expected in the next few years providing stringent tests of bound-state QED and the SME. For positronium, searches for the annual variations predicted by the SME can be performed more easily since the experiment is not accelerator based as required for muonium production. The ongoing technological developments in this field are aiming in the not too distant future to reach a precision down to a few ppt.

\section*{Acknowledgments}
The authors gratefully acknowledge the organizers of the CPT'16 conference for their kind invitation. This work has been supported by the Swiss National Science Foundation under the grant number 200020\_166286 and by the ETH Zurich Research grant ETH-35 14-1. 

\end{document}